\documentclass[11pt]{article}

\usepackage[utf8]{inputenc}
\usepackage{lipsum}
\usepackage[space]{grffile}
\usepackage{graphicx}
\usepackage{url}
\usepackage{amsmath}
\usepackage{amssymb}
\usepackage{upgreek}
\usepackage{multirow}
\usepackage{verbatim}

\usepackage{algorithm}
\floatstyle{plaintop}
\restylefloat{algorithm}
\usepackage{algpseudocode}

\usepackage{mathtools}

\usepackage{tikz}
\usepackage{pgfplots}

\usepackage{tabularx}
\usepackage{esvect}
\usepackage{array, boldline, makecell, booktabs}\newcolumntype{\$}{>{\global\let\currentrowstyle\relax}}
\newcolumntype{^}{>{\currentrowstyle}}

\newcolumntype{?}{!{\vrule width 1pt}}
\usepackage{float}  
\usepackage{tikz,tikz-3dplot}
\usetikzlibrary{calc,matrix,fit}

\usetikzlibrary{positioning,arrows.meta,quotes}
\usetikzlibrary{shapes,snakes}
\usetikzlibrary{bayesnet}
\tikzset{>=latex}

\definecolor{barBlack}{HTML}{000000}
\definecolor{barGrey}{HTML}{808080}

\begin{document}

\title{\bf Imposing Consistency Properties on\\ Blackbox Systems with Applications to\\ SVD-Based Recommender Systems}
\author{{\bf Tung D.\ Nguyen} and {\bf Jeffrey Uhlmann}\vspace{4pt} \\
Dept.\ of Electrical Engineering and Computer Science\\
University of Missouri - Columbia}
\date{}
\maketitle

\begin{abstract}
ABSTRACT: In this paper we discuss pre- and post-processing methods to induce desired consistency and/or invariance properties in blackbox systems, e.g., AI-based. We demonstrate our approach in the context of blackbox SVD-based matrix-completion methods commonly used in recommender system (RS) applications. We provide empirical results showing that enforcement of unit-consistency and shift-consistency, which have provable RS-relevant properties relating to robustness and fairness, also lead to improved performance according to generic RMSE and MAE performance metrics, irrespective of the initial chosen hyperparameter. 
\end{abstract}

\begin{footnotesize}
\begin{quote}
\end{quote}
\end{footnotesize}

\maketitle

\section{Introduction}
In this paper we examine generic transformations of inputs and outputs of a given blackbox system in order to confine the solution space of the blackbox to a subspace that is consistent with desired properties of a given application. In the context of recommender systems, an additional motivation for restricting the solution space is to limit the system's latitude to be manipulated (``gamed'') in ways that do not promote the fair and equal treatment of user information in the process of generating recommendations. More generally, restricting the space of solutions can make the system's performance more amenable to rigorous analysis. This transparency of system behavior is necessary to promote confidence in the unbiasedness of its generated recommendations, which is necessary to promote a perception of inclusivity for users from marginalized groups who may have concerns about how the information they provide may be used.  

To illustrate our general approach, we will examine the enforcement of two properties that have been proposed to be relevant to recommender system (RS) applications. The first is {\em unit consistency}. A matrix function $f(M)$ is said to be unit consistent if $f(D\cdot M\cdot E)=D\cdot f(M)\cdot E$ for positive diagonal matrices $D$ and $E$. This property can be interpreted as saying that if the units (e.g., centimeters versus meters) on input and output variables are changed, i.e., scale factors are applied to the rows and columns of $M$, then the new solution will be identical to the previous solution except given in the new units. The second property we consider is {\em shift consistency}. A matrix function $f(M)$ is said to be shift consistent if $f(M+(u\times {\mathbf 1}^{\small{T}} + {\mathbf 1}\times v^{\small{T}})))=f(M)+(u\times {\mathbf 1}^{\small{T}} + {\mathbf 1}\times v^{\small{T}})$, for column vectors $u$, $v$, and vector ${\mathbf 1}$ of all 1s. This can be interpreted as saying that if all entries in any given row or column of $M$ are shifted by a constant value, then the new solution will be identical to the previous solution but retaining the applied shifts.

In the context of recommender systems, a unit-consistent (UC) recommender system implicitly assumes that if a user Alice produces product ratings that are consistently 10\% higher than those from Bob for the same products, then if Bob were to give a new product a rating of 70/100, it is reasonable to predict that Alice is likely to give it a rating that is 10\% higher than Bob's, i.e., 77/100. This expectation is not {\em un}reasonable, especially when there is sufficient resolution in the rating system to discern scale-factor relationships like this. However, when there is low resolution, e.g., ratings are given out of a maximum of 5, then there may only be an ability to distinguish general differences in how willing different users are to give the highest (or lowest) rating. A shift-consistent (SC) recommender system, by contrast, implicitly assumes that if Alice consistently gives ratings that are 1 unit higher than those of Bob, then if Bob gives a new product a rating of 7/10, it is not {\em un}reasonable to expect Alice to give that product a rating of 8/10. 

In terms of the table/matrix of user ratings of products, UC imposes a constraint that recommendations are invariant with respect to an arbitrary positive scaling of a given row or column. SC, by contrast, is invariant with respect to the addition of a constant (i.e., a {\em shift}) to all entries of a given row or column. Our focus is on matrix-valued functions because recommender systems can be interpreted as matrix functions that transform a table/matrix of ratings of products by users\footnote{As shown in \cite{acmrs} and \cite{acmconf}, the methods we discuss in this paper can be directly generalized to tensor functions.}. Thus, imposing unit-consistency on a blackbox involves transforming its input to a unique scale-invariant canonical form, and then transforming its output back to the original problem space. Similarly, imposing shift-consistency on a blackbox involves transforming its input to a unique shift-invariant canonical form.

As illustrative examples, we will transform RS methods based on the singular value decomposition (SVD) to make them unit consistent or shift consistent. We will then examine the performance effect of imposing these constraints to test a hypothesis that eliminating extraneous degrees of freedom should reduce error according to generic metrics such as RMSE and MAE. We will then explain how such improvements can contribute to the promotion of goals such as fairness and inclusivity by reducing RS susceptibility to intentional manipulation.

\section{Canonical Transformations}

Consistent transformations are conceptually simple: {\em Find a transformation of the desired type that transforms a given matrix to a unique canonical form, then perform an operation of interest on that canonical form, and then apply the inverse transform}. The UC canonical transform performs a left and right diagonal scaling of a given nonnegative matrix such that the product of non-absent entries in each row and column is unity. The SC canonical transform, which is actually used in computing the UC form, adds a separate value to all entries in each row and column such that the sum of the shifted non-absent entries in each is zero. Absent entries can be filled so as to preserve the appropriate canonical form. For UC, replacing each absent entry with 1 preserves the unit product in its row and column, and SC just requires replacing absent entries with 0 to preserve the sum (zero) of its row and column.

The well-known singular value decomposition (SVD) defines a transform that is consistent with respect to left and right orthonormal / unitary transformations, where the diagonal matrix of sorted nonnegative singular values is the unique canonical form\footnote{We will sometimes use ``SVD'' to refer to the decomposition and sometimes to the canonical diagonal matrix of singular values when the context provides clear disambiguation.}. Thus, for $f(M)$ giving the singular value canonical form of $M$, $f(U\cdot M\cdot V) = U\cdot f(M)\cdot V$, where $U$ and $V$ are orthonormal matrices. The SVD also suggests a natural utility function for matrix completion: {\em jointly fill the missing entries so as to minimize the rank, i.e., the number of nonzero singular values, of the completed matrix}. This can be interpreted as completing the matrix in way that minimizes the spurious incorporation of extra structure / information into the result. As an intuitive example, if every filled entry of the matrix is 1, then replacing all absent entries with 1 gives the minimum-rank solution. This solution makes intuitive sense because there is no information available that could justify replacing a given absent entry with any value other than 1.

\section{SVD-Based Methods}

The seminal use of SVD-based matrix completion was given in \cite{firstSVD}. Many variants have subsequently been proposed and evaluated, where in most cases the principal differences are focused on reducing the $O(n^3)$ computational cost of evaluating the full decomposition of an $n\times n$ matrix\footnote{The complexity to evaluate the SVD for a general $m\times n$ matrix is $O(mn\cdot \min(m,n)))$, but we will consider the $n\times n$ case purely for ease of interpreting and comparing the relative computational complexities of different methods.}. More specifically, most methods retain only the $k$ largest singular values so that the overall SVD complexity becomes $O(kn^2)$, which for small $k$ approaches the optimal $O(n^2)$ complexity of the UC and SC methods. Thus, for these methods the choice of $k$ becomes a hyperparameter, and the manner of choosing its value is what distinguishes them\footnote{For our purposes we will focus on a ``typical'' representative approach with no attempt to capture the full breadth of literature on SVD-based completion methods.}.

\begin{figure*}
	\centering
		\includegraphics[scale=0.45]{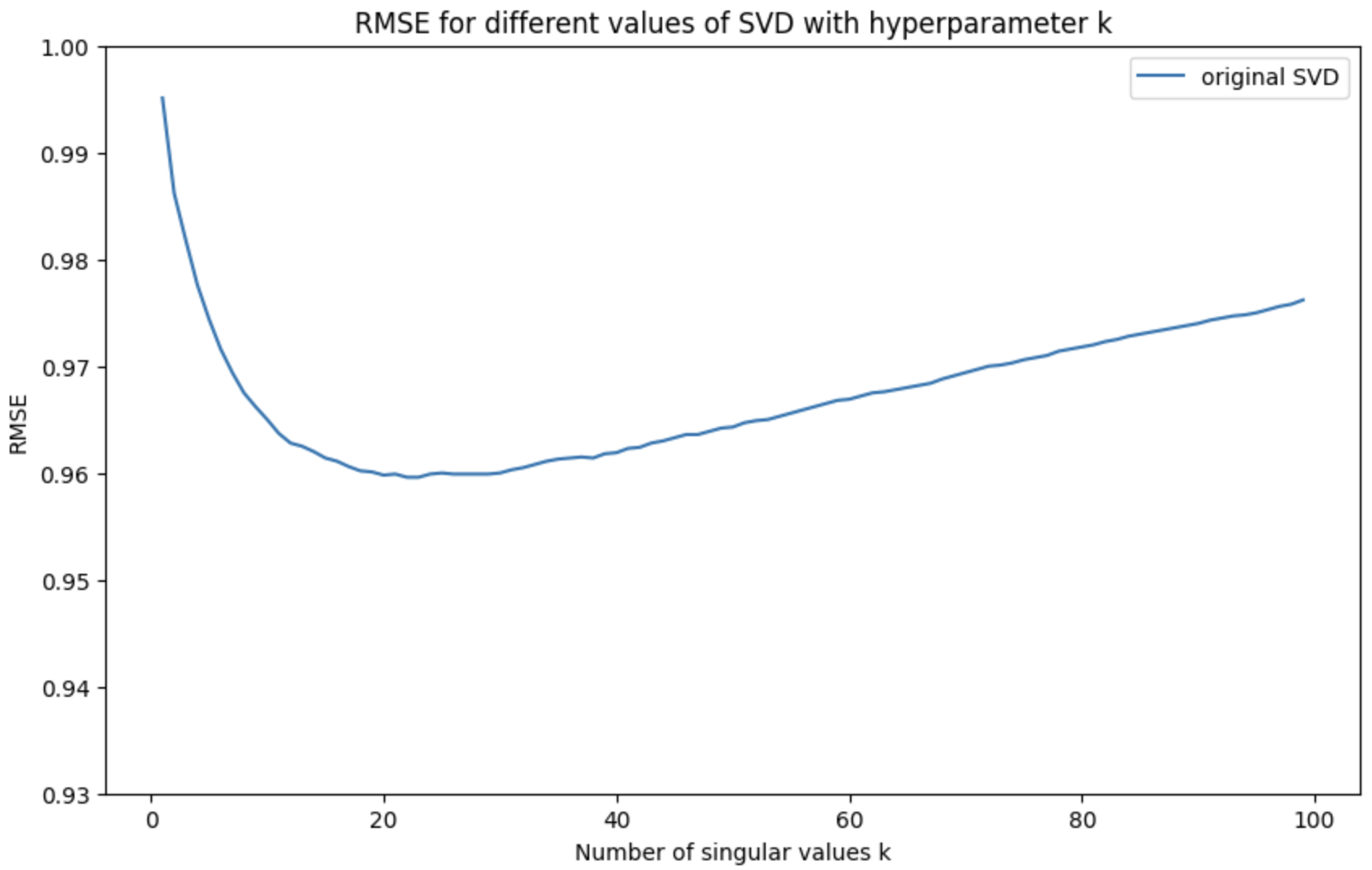}
	\caption{\footnotesize The graph shows the RMSE of predicted ratings/recommendations for each choice of value for the hyperparameter $k$ of retained singular values. As can be seen, there is an approximate minimum in the neighborhood of $k=25$.}
	\label{kgraph}
\end{figure*}

The most agnostic approach for selecting the value of $k$ is to achieve a specified performance threshold. An alternative approach is to determine its value empirically in a way that makes a tradeoff between the amount of information provided by the $k$ retained singular values versus the computational cost to retain that number of singular values. One intuitive assumption is that performance according to a given measure, e.g., RMSE or MAE, will suffer if $k$ is too small because important application-specific information will not be retained and exploited. Another intuitive assumption is that many of the small singular values will represent ``noise'' that is not relevant to the application and thus may tend to degrade performance. It may thus be hypothesized that that a graph of choices of $k$ will exhibit a minimum somewhere between the two extremes. As shown in Figure \ref{kgraph}, such a minimum often exists (at least approximately). 

A judicious choice of $k$ based on a generic performance measure is often effective for improving performance according to that metric, and it is not unusual for a given choice to optimize many different metrics. However, any choice of $k$ less than the rank of the matrix will almost certainly discard some amount of relevant application-specific information and retain some amount of corrupting noise. Thus, while SVD-based methods can exhibit good performance according to generic metrics, very little can be rigorously stated about its actual performance properties, e.g., the extent to which discarded singular values may correspond to preference information relating to minority sub-populations within the user base. It also prevents the satisfaction of the consensus order criterion proposed in \cite{acmrs}. A related concern is the extent to which the hyperparameter $k$ may permit users to unfairly increase the weight given to their set of relative preferences/ratings by uniformly scaling or shifting their set of ratings in a manner that preserves their rank order but increases their influence on system recommendations to other users.

One way to limit the sensitivity of a blackbox SVD-based method is to externally impose a consistency condition on its performance by transforming its input to a scale-invariant or shift-invariant canonical form, obtain its results in that invariant space, and then transform the results back to the original space\footnote{In the unit-consistent case, this is equivalent to using the scale-invariant SVD defined in \cite{siam-jku}.}. It is hypothesized that imposing constraints that are {\em meaningful} with respect to the application will have two beneficial effects:
\begin{enumerate}
    \item The imposed constraint will eliminate a degree of freedom that is potentially available for users to manipulate their ratings to increase their influence on recommendations to other users.
    \item The imposed constraint will tend to improve the performance of the system by narrowing the available solution space by restricting variables that do not necessarily relate to desired system behaviors, e.g., preventing a user from influencing the rank order of recommendations by uniformly scaling or shifting their ratings to higher ones.
\end{enumerate}
In the next section we empirically test these hypotheses. 

\section{UC/SC Effects on SVD RS Performance}

In this section we provide test results examining the effects of imposing UC and SC constraints on the quality, in terms of RMSE and MAE, of SVD recommendations using the MovieLens1M and MovieLens100k dataset \cite{Movielens100k+1M}. 

\begin{figure*}
	\centering
		\includegraphics[scale=0.4]{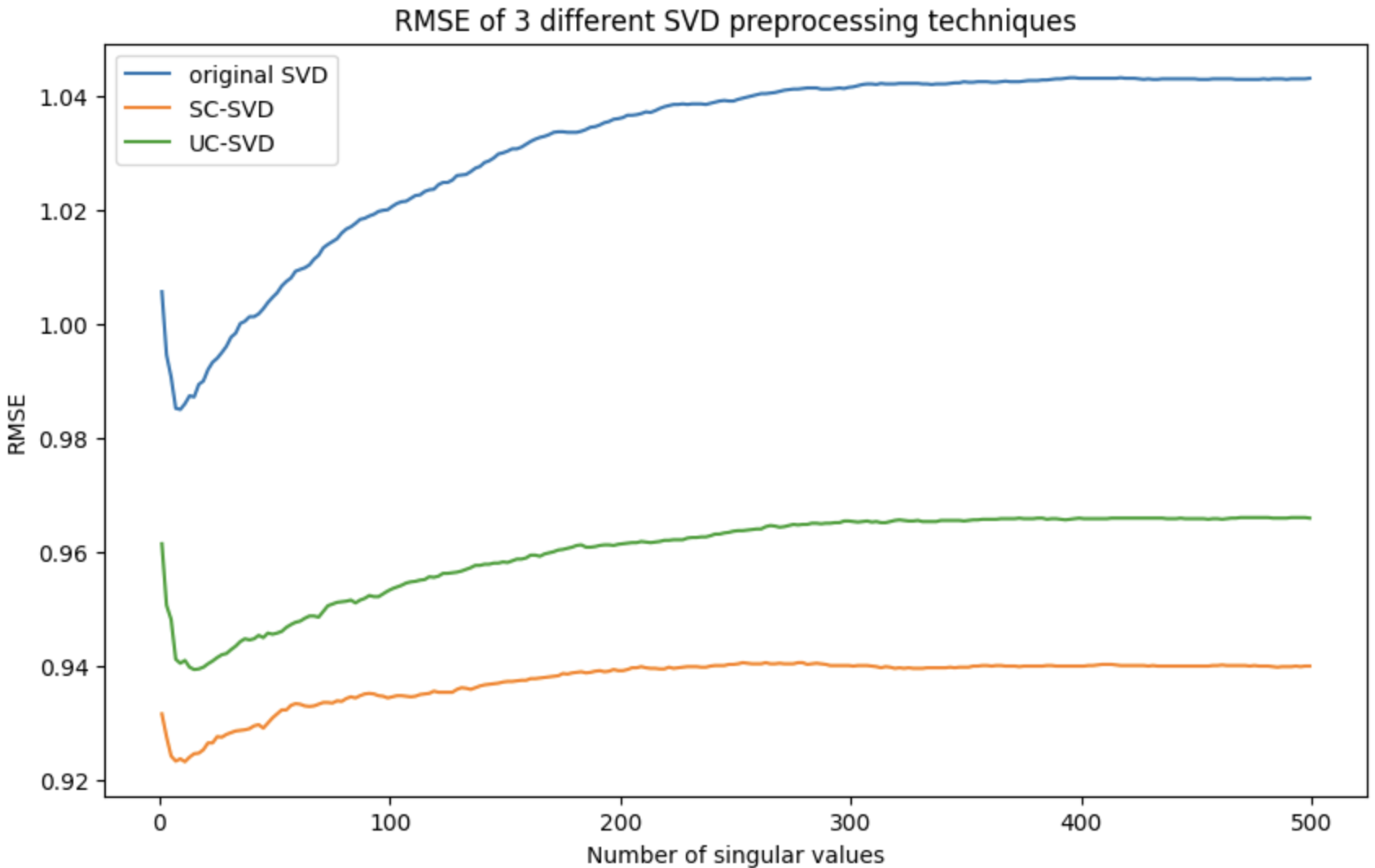}
	\caption{\footnotesize The graph shows the RMSE of predicted ratings/recommendations for each UC-SVD and SC-SVD in comparison with original SVD in Movielens 100k.}
	\label{ucscsvd-rmse}
\end{figure*}

Figure \ref{ucscsvd-rmse} shows the RMSE performance of the original SVD method, and its performance when UC-constrained and SC-constrained. As can be seen, both constrained variants exhibit improved performance, with the SC constraint providing slightly smaller RMSE than the UC constraint. 

\begin{figure*}
	\centering
		\includegraphics[scale=0.4]{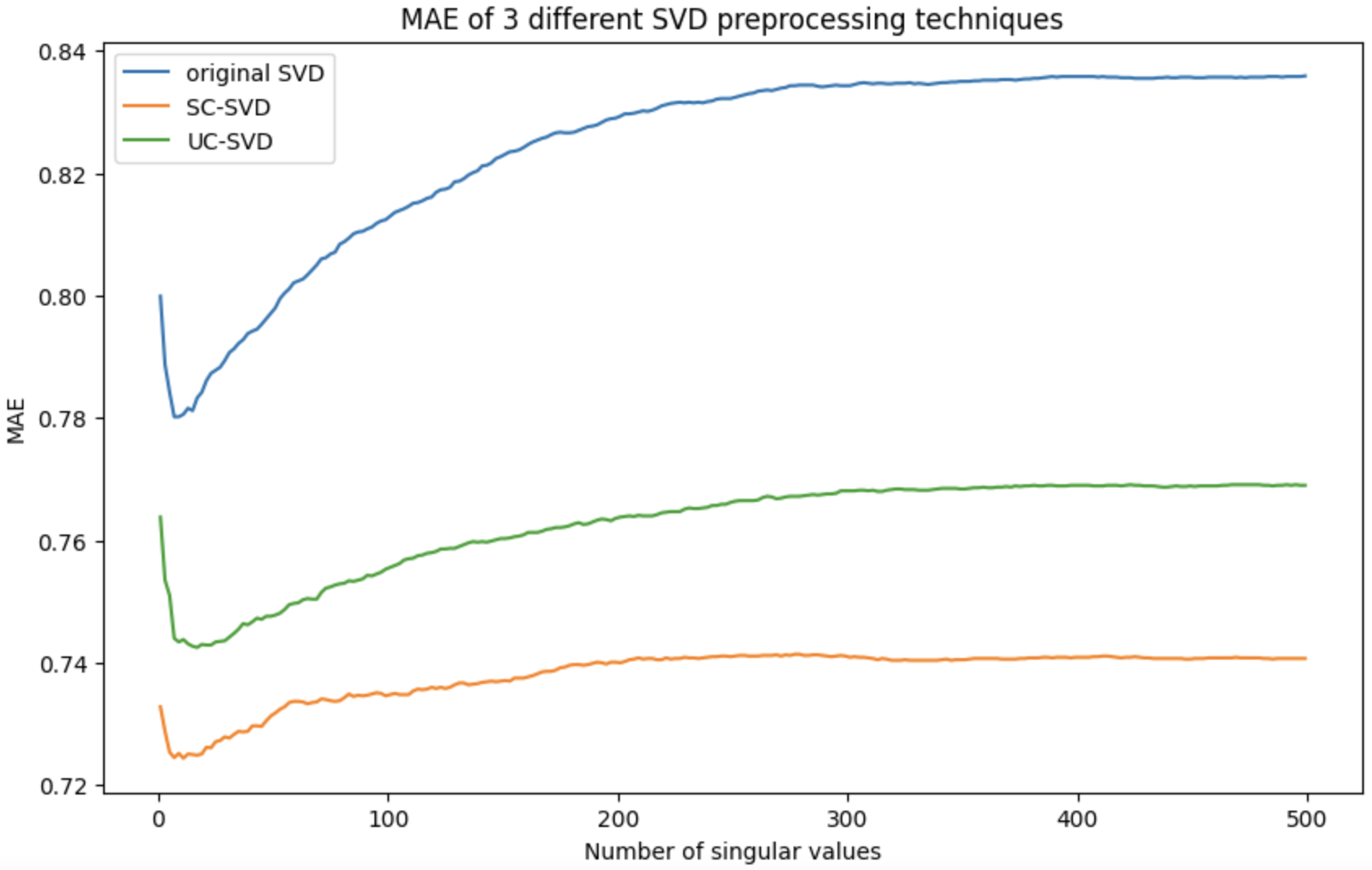}
	\caption{\footnotesize The graph shows the MAE of predicted ratings/recommendations for each UC-SVD and SC-SVD in comparison with original SVD in Movielens 100k.}
	\label{ucscsvd-mae}
\end{figure*}

Figure \ref{ucscsvd-mae} shows qualitative similar relative behaviors according to MAE.  It has been suggested in \cite{acmrs, acmconf} that UC may be more sensitive to discretization effects resulting from ratings being restricted to a small number of integer values, e.g., 1 to 5 in the MovieLens dataset used for these tests. The results shown in Figures \ref{ucscsvd-rmse} and \ref{ucscsvd-mae} may provide evidence supporting that hypothesis. 

\begin{figure*}
	\centering
		\includegraphics[scale=0.4]{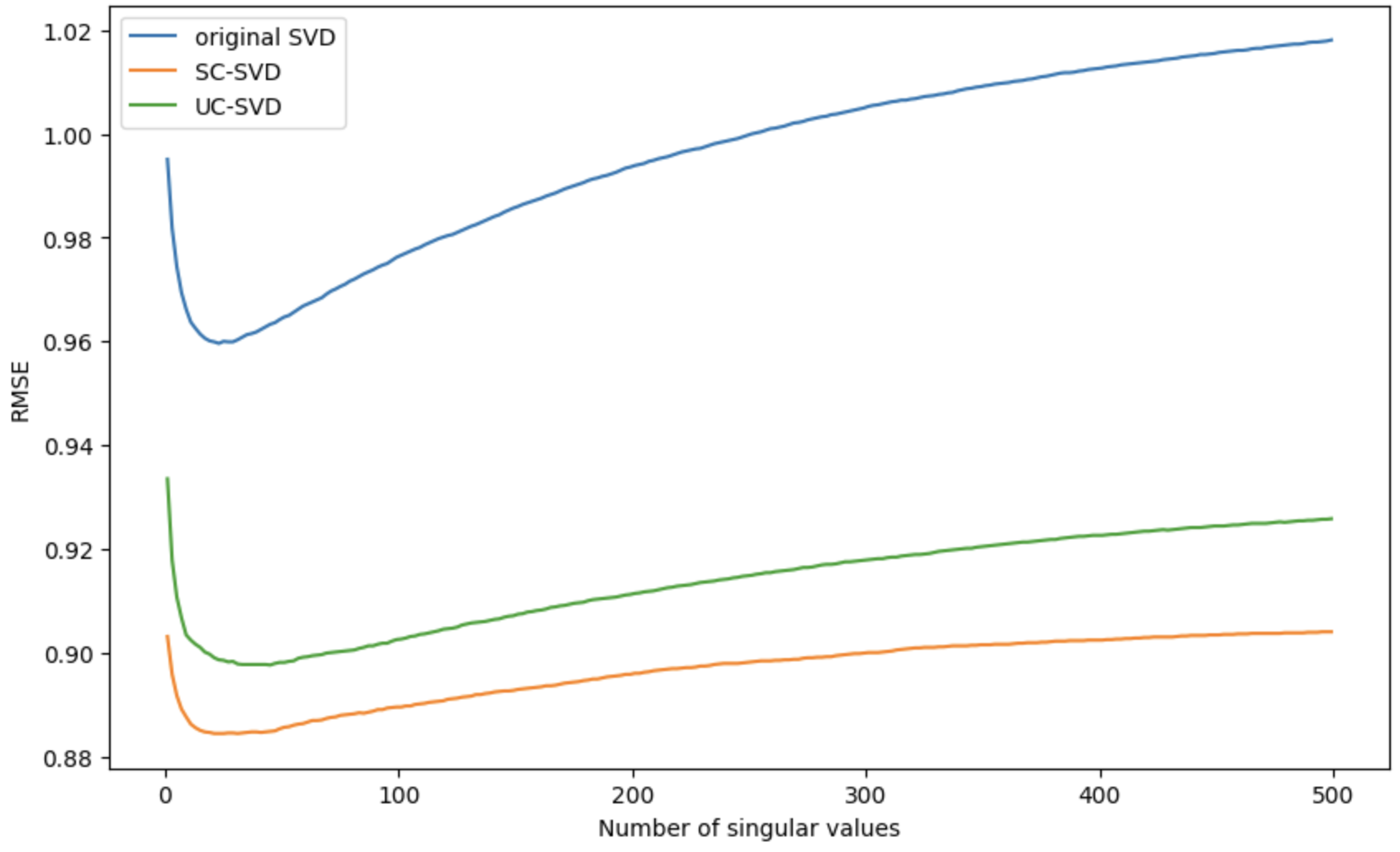}
	\caption{\footnotesize The graph shows the RMSE of predicted ratings/recommendations for each UC-SVD and SC-SVD in comparison with original SVD in Movielens 1M.}
	\label{mae-ml1m}
\end{figure*}

\begin{figure*}
	\centering
		\includegraphics[scale=0.4]{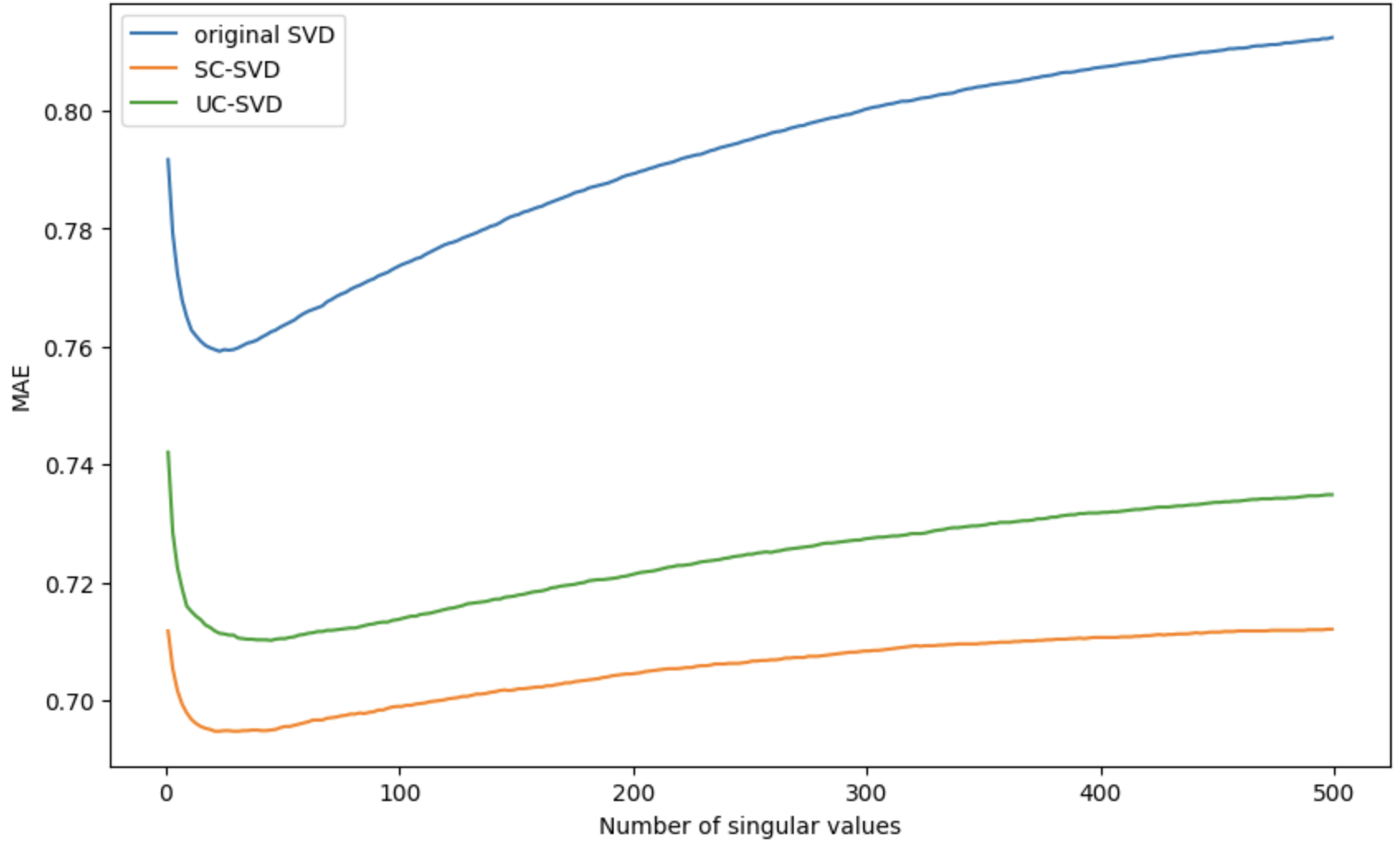}
	\caption{\footnotesize The graph shows the MAE of predicted ratings/recommendations for each UC-SVD and SC-SVD in comparison with original SVD in Movielens 1M.}
	\label{rmse-ml1m}
\end{figure*}

As discussed in \cite{acmrs, acmconf}, both unit-consistency and shift-consistency tend to limit the extent that an individual user can increase the influence of their relative preferences on system recommendations given to other users. In the case of an SVD-based RS, for example, if Alice were to scale her ratings so that her highest given rating is equal to the highest possible rating, or if she were to achieve the same by incrementing all of her ratings, the increased magnitude of her vector (row/column) of ratings would result in a commensurately increased weight given by the SVD to her relative preferences. Such manipulation by Alice would clearly be ``unfair,'' and our results suggest that mitigating it by imposing either a UC or SC constraint not only increases fairness, it also appears to improve performance by eliminating noise/error from the eliminated degree of freedom.

\section{Summary}

In this paper we have briefly introduced and examined the imposition of constraints on a blackbox SVD-based recommender system. We have also provided empirical evidence supporting hypotheses relating to how such constraints may lead to improved performance according to standard measures such as RMSE and MAE. These positive results should motivate future examinations of how different constraints can enhance the performance of blackbox systems in other problem domains. 

It must be emphasized that relative performance differences according to generic metrics, e.g., RMSE and MAE, are not necessarily meaningful in and of themselves. If improvements derive from the imposition of constraints that are deemed relevant to the problem domain, then it is reasonable to conclude they are a consequence of reducing noise/variation from degrees of freedom that are restricted by those constraints. The fact that extraneous degrees of freedom may introduce noise into the recommendation process should not be surprising, and it is tempting to conclude that the magnitude of the performance improvements obtained by restricting them are a measure of their significance. However, while the effect may be small in the case of random noise, the magnitude of effect could be vastly larger under the influence of non-random, i.e., active (possibly nefarious), manipulation. In other words, the reason to restrict degrees of freedom to only those relevant to the problem domain is to ensure that they cannot be later exploited to manipulate recommendations in undesired ways.    

Clearly there is no way to make a system invulnerable to corruption from invalid inputs, i.e., garbage-in/garbage-out, but it is possible to constrain opportunities for users to corrupt the system in ways that achieve specific benefits to them.  UC and SC constraints promote fairness by limiting the extent to which users can influence the system to give more weight to their relative preferences when generating recommendations to other users. More generally, such constraints tend to make the system easier to analyze, e.g., amenable to proving rigorous performance properties, and such transparency tends to promote trust in the system. Fairness and trust are essential qualities for attracting a diverse and inclusive user base, especially in the case of a recommender system that inherently requires users to provide potentially sensitive information about their personal preferences.

\end{document}